# An Approximate Expression for Viscosity of Nanosuspensions

N.G.Domostroeva[1] and N.N.Trunov[2]

*D.I.Mendeleyev Institute for Metrology*

*Russia, St.Peterburg. 190005 Moskovsky pr. 19*

December, 11, 2009

**Abstract:** We consider liquid suspensions with dispersed nanoparticles. Using two-points Pade approximants and combining results of both hydrodynamic and molecular dynamics methods, we obtain the effective viscosity for any diameters of nanoparticles.

## 1. Introduction

Suspensions with the dispersed micro- and nanoparticles are currently of great interest since they are used or will be used in many nanotechnologies. A very important part of them is connected with oil and gas industry [1,2].

For many practical rheological measurements it is necessary to have simple convenient formulas for viscosity of such nanonsuspensions in a wide range of theirs diameters, masses and concentrations. Meanwhile, consistent results exist only in the hydrodynamic limit for small concentrations.

In the present paper we have made an attempt to construct such a formula using general physical considerations and only minimum of calculated data.

---

[1] Electronic address: N.G.Domostroeva@vniim.ru
[2] Electronic address: trunov@vniim.ru

## 2. The hydrodynamic approach

The following formula for the effective viscosity $\eta$ of a rarefied coarse-dispersed suspension is well known:

$$\eta = \eta_0 (1 + 2.5\varphi), \qquad (1)$$

where $\eta_0$ is the viscosity of the carrier liquid and $\varphi$ is the volume concentration of dispersed particles. Since the equation (1) is only valid for very small $\varphi$ we'll use two-points Pade approximants in order to extrapolate (1) for larger values of $\varphi$. Such approximants have the form

$$\Pi_{km}(\varphi) = \frac{A_k \varphi^k + A_{k-1} \varphi^{k-1} + \ldots + A_0}{B_m \varphi^m + B_{m-1} \varphi^{m-1} + \ldots + B_0} . \qquad (2)$$

The best result is often achieved if $k = m$. We use the simplest form $k = m = 1$ which requires minimum of preliminary information.

Let us consider a gas of spherical dispersed particles as if they were in vacuum. It is known that the plot of the pressure $P$ as a function of $\varphi$ may be continued up to $\varphi = \varphi^* = 0.5$; in this range the phase transition to a liquid takes place. Liquid metals usually have $\varphi = 0.45\ldots0.47$.

It is naturally to suppose that the viscosity of such suspension must tend to infinity when $\varphi \to \varphi^*$ since greater values of $\varphi$ cannot be continuously reached. We can take into account this supposition by writing:

$$\eta = \eta_0 \frac{1 + a\varphi}{1 - \varphi/\varphi^*} , \qquad (3)$$

where the value $a$ must be chosen so that (3) coincides with (1) in linear approximation:

$$\eta/\eta_0 = \frac{1 + (2.5 - b)\varphi}{1 - b\varphi} ; \quad b = \frac{1}{\varphi^*} . \qquad (4)$$

Expanding (4) into the series we get at small $\varphi$

$$\eta/\eta_0 = 1 + 2.5\varphi + 2.5b\varphi^2 . \qquad (5)$$

Assuming $\varphi^* = 0.5$ so that $b = 2$ we obtain the quadratic term $5\varphi^2$; an improvement of the hydrodynamic calculation leads to the same value [3].

Alternative calculations have the quadratic term in the interval between $5\varphi^2$ and $6.25\varphi^2$. In the last case the Pade form (4) is especially simple: $b = 2.5$, $\varphi^* = 0.4$, so that

$$\eta/\eta_0 = \frac{1}{1 - 2.5\varphi} \qquad (6)$$

Remember that the linear approximation (1) and its improvements like (5) are obtained within the frame of usual hydrodynamics so that one of the necessary conditions for their validity is

$$\delta = d/D \ll 1 , \qquad (7)$$

where $d$ is diameter of particle of the carrier liquid and $D$ is diameter of the dispersed particle.

On the contrary, the value of $\varphi^*$ is pure geometrical by origin. Thus one may expect that the denominators as in (3) or (4) retain their sense independently on the value $\delta$ and other parameters.

# 3. An approximate formula for arbitrary dispersed particles

In the case of relatively small micro- and nanoparticles the hydrodynamic approach treating the carrier liquid as a homogeneous one is inapplicable. Some separate results are obtained when the effective viscosity was simulated using the molecular dynamics methods [4]. It is stated there that essential parameters are not only $\varphi$, but also $\delta$ (7) and

$$\mu = M/m , \qquad (8)$$

where $M$ is the mass of nanoparticle and $m$ is the mass of the fluid molecule.

The ratio $\mu$ is obviously connected with each collision "nanoparticle-molecule". The number of such collisions is proportional to $\varphi$, so that it is naturally to expect that $\eta$ depends on $\mu$ by means of the product $\mu\varphi$.

According to [4] there is a linear dependence of the effective viscosity upon the ratio $r$ of the density of the nanoparticles material to the carrier liquid density. Since $r$ is proportional to $\mu\delta^3$, we can write instead of (4):

$$\eta/\eta_0 = \frac{1+(2.5-b+qr)\varphi}{1-b\varphi} \qquad (9)$$

with a constant $q$.

The ratio of these densities may be written as

$$r = \frac{M}{\pi D^3/6} : \frac{m}{\alpha d^3/\sqrt{2}} = 1.35\beta \; ; \qquad (10)$$

$$\beta = \mu\delta^3,$$

$\alpha = 1$ corresponds to the close packing of the carrier liquid molecules.

Hereafter we use the value of the linear term $0.16r$ given in [4] for $\alpha = 5$ and $\varphi = 0.129$.

For the most probable values $\varphi^* = 0.5$, $b = 2$, substituting given values of $\alpha$ and $\varphi$ into (9), (10), we obtain

$$\eta/\eta_0 = \frac{(1+0.5+1.25\beta)\varphi}{1-2\varphi} \quad . \qquad (11)$$

In the second limiting case $b = 2.5$

$$\eta/\eta_0 = \frac{1+1.13\beta\varphi}{1-2.5\varphi} \quad . \qquad (12)$$

We could also take into account more complicated dependence of $\eta$ on $\beta$ and $\varphi$, but this needs additional detailed and systematic data.

 **Acknowledgment**
Authors are grateful to Dr. A.A.Belkin for his explanations of the text [4].